\documentclass[english]{revtex4-1}
\usepackage[T1]{fontenc}
\usepackage[latin9]{inputenc}
\usepackage{amsmath}
\usepackage{amsthm}
\usepackage{graphicx}
\PassOptionsToPackage{normalem}{ulem}
\usepackage{ulem}

\makeatletter

\newcommand*\LyXZeroWidthSpace{\hspace{0pt}}
\providecommand{\tabularnewline}{\\}

\numberwithin{equation}{section}
\numberwithin{figure}{section}

\@ifundefined{date}{}{\date{}}
\@ifundefined{showcaptionsetup}{}{%
 \PassOptionsToPackage{caption=false}{subfig}}
\usepackage{subfig}
\makeatother

\usepackage{babel}
\begin{document}

\title{Determination of the strong coupling constant from ATLAS measurements
of the inclusive isolated prompt photon cross section at $\sqrt{s}=7$
TeV.}

\author{Boussaha Bouzid}
\email{bo.boussaha@gmail.com}

\author{Farida Iddir}
\email{iddir.farida@univ-oran.dz}

\author{Lahouari Semlala}
\email{semlala.lahouari@univ-oran.dz}

\address{Laboratoire de Physique Théorique d'Oran (LPTO), University of Oran1-Ahmed
Ben Bella; BP 1524 El M\textquoteright nouar Oran 31000, ALGERIA.}
\begin{abstract}
We present an estimation of the strong coupling constant $\alpha_{s}(M_{Z}^{2})$
using, for the first time, the production of prompt photon process
in proton-proton collisions at the LHC. ATLAS measurements of the
inclusive isolated prompt photon cross section at $\sqrt{s}=7$ TeV
are exploited. Both theoretical and experimental uncertainties are
estimated and the strong coupling constant has been determined to
be $\alpha_{s}(M_{Z}^{2})=0.1183\pm0.0038,$ to NLO accuracy. 
\end{abstract}
\maketitle

\section{Introduction}

The strong coupling \textquotedbl{}constant\textquotedbl{} $\alpha_{s}$
is the basic free parameter of Quantum Chromodynamics. QCD predicts
that $\alpha_{s}(Q^{2})$ decreases with increasing energy or momentum
transfer $Q$, and vanishes at asymptotically high energies. Testing
the energy dependence (running) of $\alpha_{s}$ over a wide range
provides an implicit test of QCD. Any modified running of the strong
coupling may be a sign of new physics, for instance, a possible existence
of new coloured matter near TeV energies is considered in Ref.  \cite{new coloured matter}. 

$\alpha_{s}$ is not an ``observable'' by itself. Values of $\alpha_{s}(Q^{2})$
are determined from measurements of observables for which QCD predictions
exist. Different particle reactions and scattering processes, performed
at different energy scales $Q^{2}$, are used to extract the strong
coupling parameter$^{[2]}$. In the theoretical framework, the Lattice
QCD uses several approaches ``which directly determine $\alpha_{s}$
on the lattice in a scheme closer to $\overline{\mathrm{MS}}$''$^{[3]}$;
the ETM Collaboration uses a comparison of lattice data for the ghost-gluon
coupling with that of perturbation theory$^{[4]}$, providing the
first determination of $\alpha_{s}$ with 2+1+1 flavors of dynamical
quarks.

The energy reach available at the LHC makes possible, for the first
time, to perform direct ``measurements'' of $\alpha_{s}(Q^{2})$
in the TeV scale. Jet production is used to extract $\alpha_{s}$
and to test its running with the momentum transfer up to the TeV region,
currently to NLO accuracy$^{[5]}$and the CMS Collaboration reported
the first determination of $\alpha_{s}$ using events from top-quark
production, to NNLO accuracy$^{[6]}$.

Here, we propose for the first time an extraction of the strong coupling
constant using the prompt photon production process at the LHC. ATLAS
measurements of the inclusive isolated cross section at $\sqrt{s}=7$
TeV is presented as a function of transverse energy $E_{T}^{\gamma}$
of the photon in the kinematic range $15\le E_{T}^{\gamma}<1000$
GeV and the pseudo rapidity $\eta^{\gamma}$ regions $|\eta^{\gamma}|<1.37$
and $1.52\le|\eta^{\gamma}|<2.37$$^{[7,8,9]}$. NLO calculations
are performed using JETPHOX $^{[10]}$ with (NLO) CT10w parton density
functions$^{[11]}$, provided by the LHAPDF package$^{[12]}.$

The paper is organized as follows. In Sec. \ref{sec:The-theoretical-prediction}
we present the theoretical predictions, involving the perturbative
part and the non-pertubative corrections. Section \ref{sec:Extraction-of-}
describes the extraction of the $\alpha_{s}$ from data and the averaging
procedure. Both scale and PDF uncertainties are considered. The conclusion
is presented in Sec. \ref{sec:Conclusion}. 

\section{\label{sec:The-theoretical-prediction}The theoretical predictions}

NLO differential cross-sections of isolated prompt photon production
are calculated with the JETPHOX program$^{[10]}$. Further details
on the scale and pdf uncertainties are available in papers \cite{note CERN,note ATL},
in region of phase space of interest.

The radius of the isolated cone is set to $R=0.4$ in $\eta-\varphi$
space around the photon direction and the maximum transverse energy
cut deposited in the isolation cone (at the parton-level) is set to
$(E_{T}^{\mathrm{iso}})_{max}=4\:\mathrm{and\:7\,GeV}$ as recommended
in papers \cite{ATLAS1} and \cite{ATLAS3} respectively. Note that
all ATLAS and CMS prompt photon measurements use the same definition
of the cone isolation variable $E_{T}^{\mathrm{iso}}$ with a unique
cone radius value $R=0.4$. This value seems to be the most suitable
for the analysis.

The theoretical prediction are multiplied by an additional correction
factor $C_{\mathrm{np}}$ to account for the presence of contributions
from the underlying event and hadronization, using the Monte Carlo
generator PYTHIA$^{[15]}$.

The renormalization $\mu_{R}$, factorization $\mu_{F}$ and fragmentation
$\mu_{f}$ scales are set to be equal: 
\begin{equation}
\mu_{R}=\mu_{F}=\mu_{f}=\mu=E_{T}^{\gamma},\label{scales-1}
\end{equation}

and the scale effects are evaluated using the uncertainty band varying
the scales coherently and independently, (see Sec. \ref{subsec:The-scale-effect}).

\subsubsection*{NLO calculations}

The knowledge of the $\left(d\sigma/dE_{T}^{^{\gamma}}\right)_{\mathrm{NLO}}$
as functions of $\alpha_{s}(M_{Z}^{2})$ in each $(|\eta^{\gamma}|\mathrm{\mathrm{-}}E_{T}^{\gamma})$
bin is allowed using \texttt{CT10wnlo\_as\_xxxx} parametrizations
presented for sixteen $\alpha_{s}(M_{Z}^{2})$ values in the range
$0.112\mathrm{-}0.127$ in steps of $0.001$.
\begin{figure}
\subfloat[$|\eta^{\gamma}|<0.6$]{\includegraphics[scale=0.33]{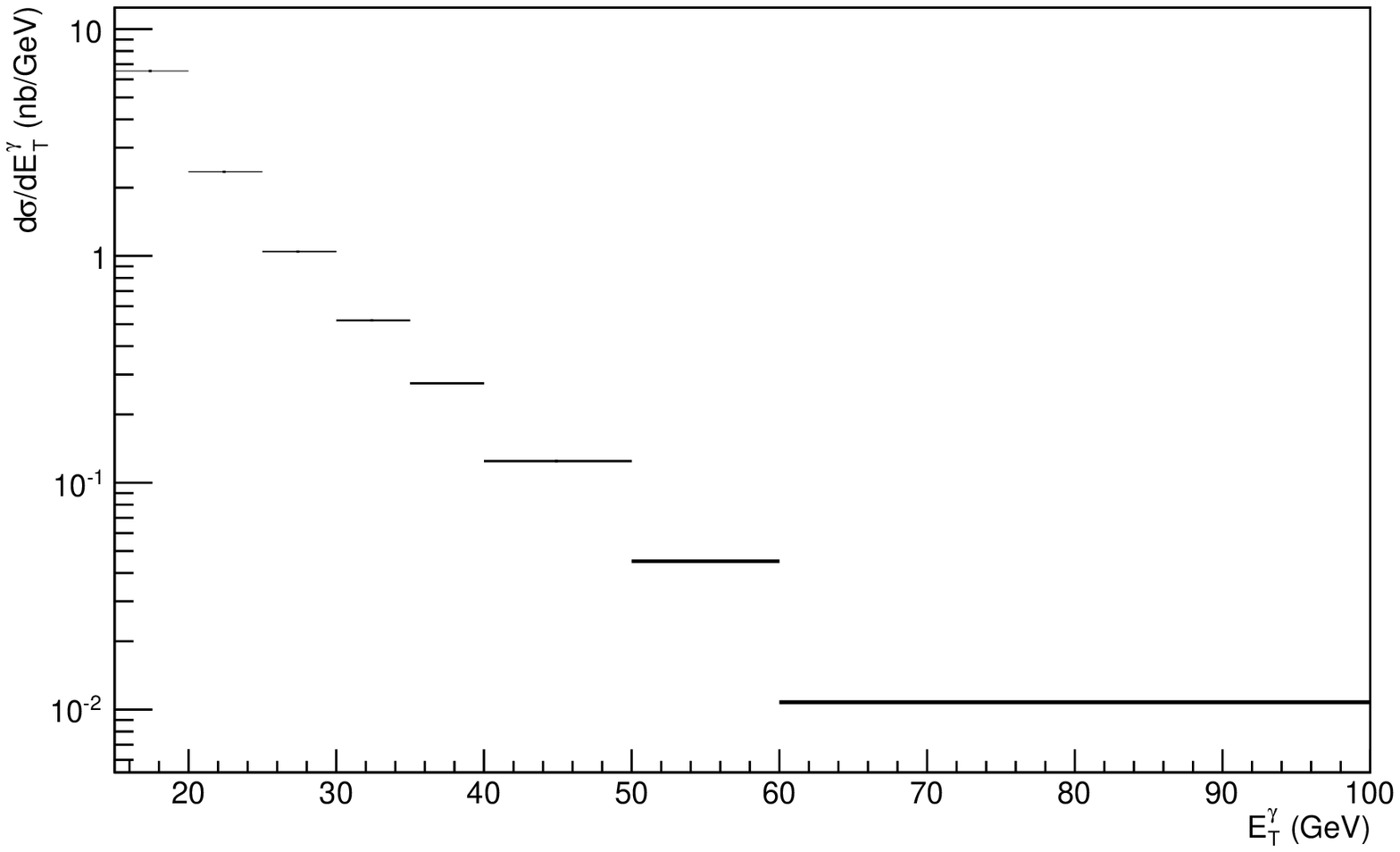}}\subfloat[$0.6<|\eta^{\gamma}|<1.37$]{\includegraphics[scale=0.33]{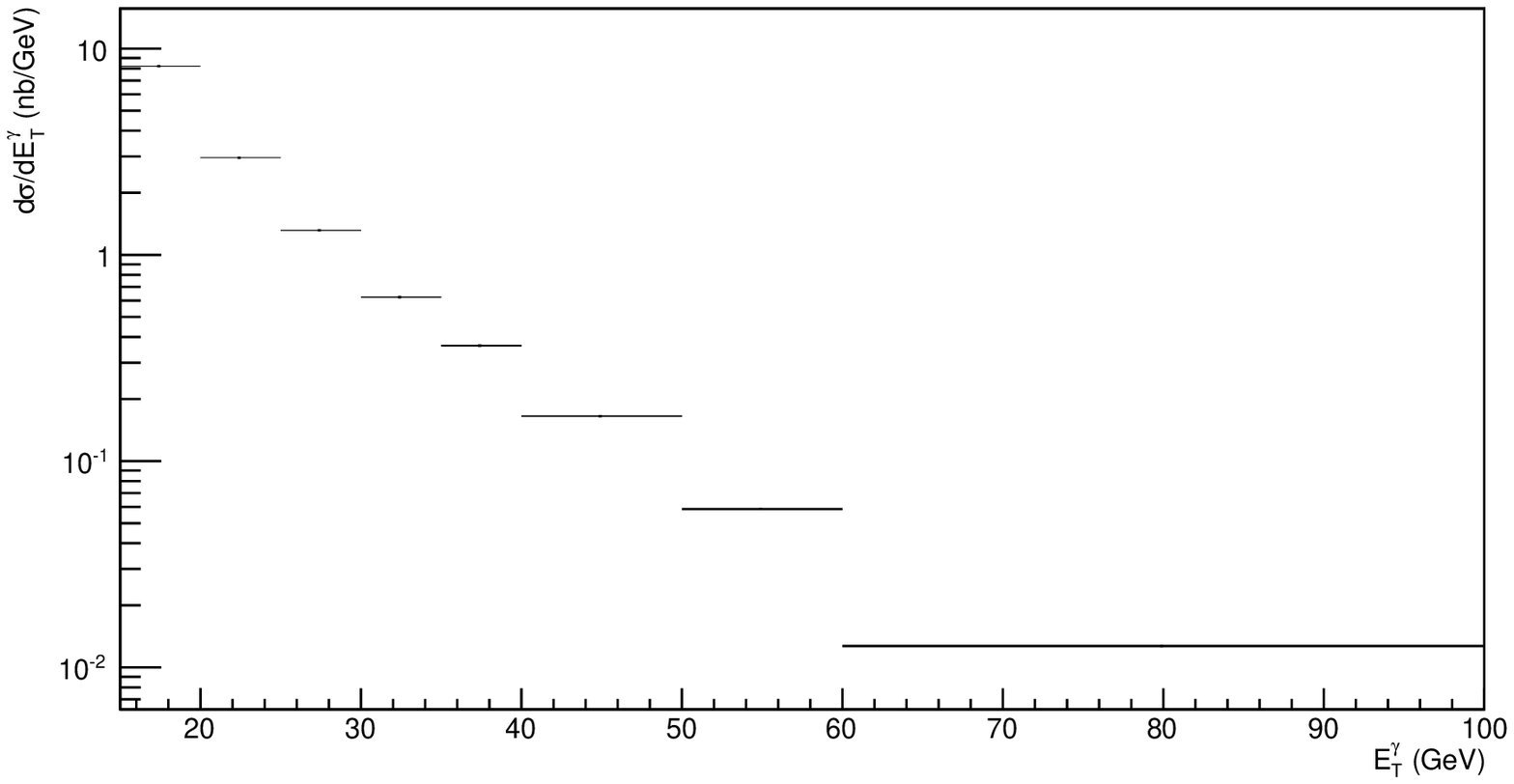}}\linebreak{}
\subfloat[ $1.52<|\eta^{\gamma}|<1.81$]{\includegraphics[scale=0.41]{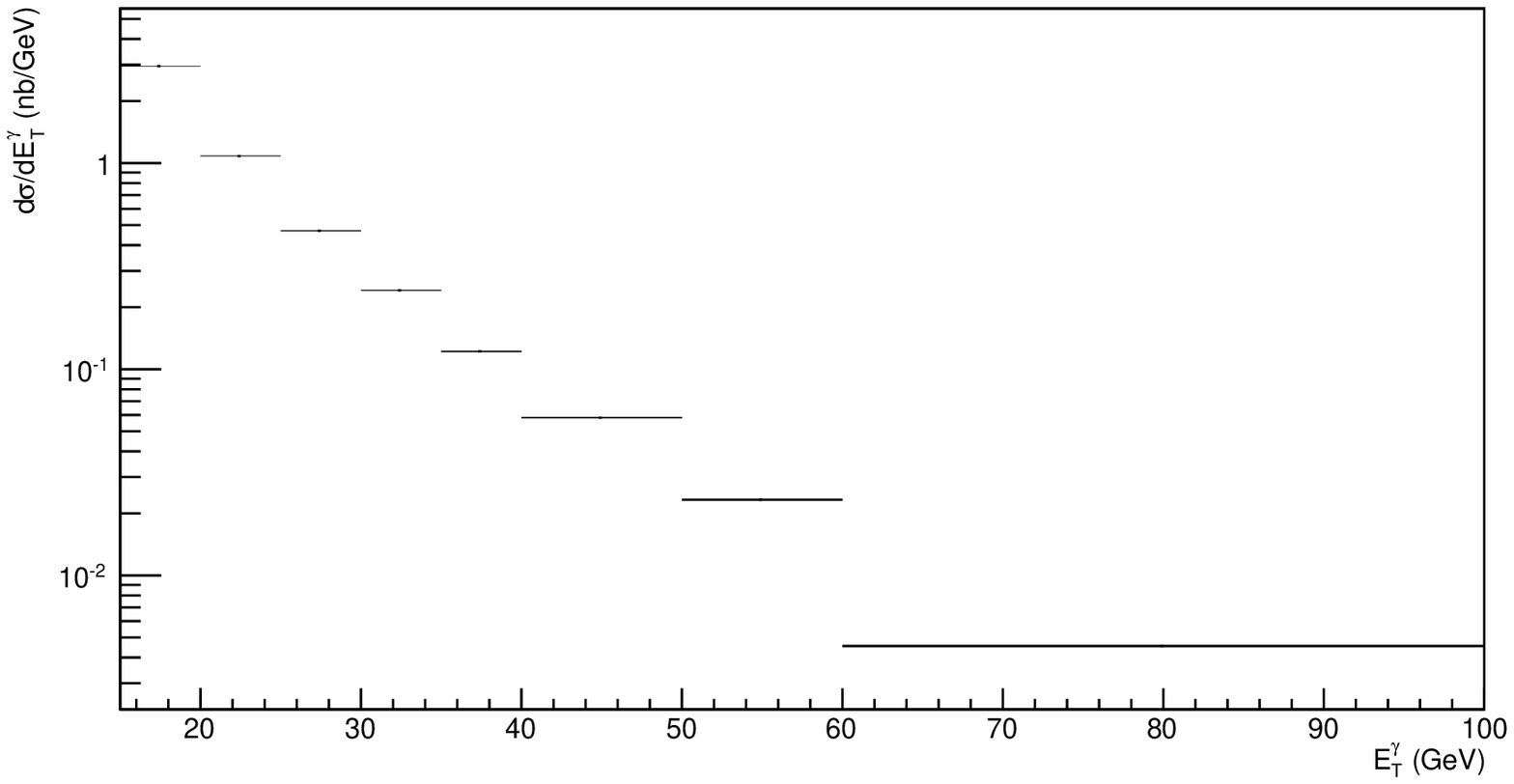}}\caption{\label{fig:NLO-differential-cross-sections}NLO differential cross-sections
multiplied by the non-prturbative coefficient $C_{\mathrm{np}}$,
(\texttt{CT10wnlo\_as\_0118}, $E_{\mathrm{iso}}<4$ GeV) }
\end{figure}

The choice of this CTEQ set is motivated by its fine alphas scan and
by the fact that it is largely used in theoretical calculations with
consistent results in comparisons with data. 

JETPHOX calculations are done for each pdf member (i.e. for each value
of $\alpha_{s}(M_{Z}^{2})$), an example is shown in Fig.\ref{fig:NLO-differential-cross-sections}.
NLO calculations yield a set of cross section values related to their
corresponding $\alpha_{s}$. For each $(|\eta^{\gamma}|\mathrm{-}E_{T}^{\gamma})$
bin we construct a one-to-one mapping, noted $f_{\mathrm{bin}}$,
between the calculated cross section $\frac{d\sigma_{i}}{dE_{T}^{^{\gamma}}}$,
related to the pdf member $i$ (16 pdf members), and its corresponding
strong coupling $\alpha_{i}$:

\begin{equation}
\alpha_{i}=f_{\mathrm{bin}}\left[\left(\frac{d\sigma_{i}}{dE_{T}^{^{\gamma}}}\right)_{\mathrm{NLO}}\right],\label{ascurve}
\end{equation}

with 
\begin{equation}
\mathit{i=\mathrm{0,...15;}}\quad\alpha_{i}=\mathit{\mathrm{0.112}+i*\mathrm{0.001.}}\label{ascurve-1}
\end{equation}

\begin{figure}
\includegraphics[scale=0.43]{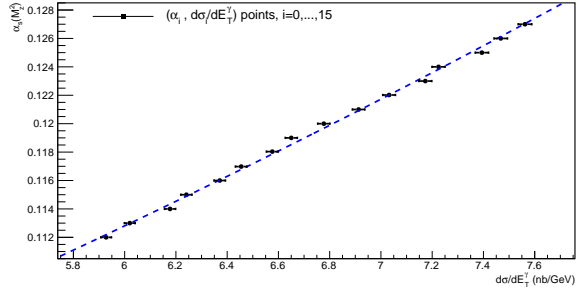}\caption{\label{alphas_vs_ds}$\alpha_{s}\left(M_{Z}^{2}\right)$versus $d\sigma/dE_{T}^{\gamma}$,
in the range $15<E_{T}^{\gamma}<20$ GeV and $|\eta^{\gamma}|<0.6$.}
\end{figure}

We have 79 maps corresponding to all possible bins needed for our
analysis, one such curves is illustrated in Fig.\ref{alphas_vs_ds}.

Note that JETPHOX calculations use a running $\alpha_{s}(Q^{2})$
extracted from the LHAPDF routine, assuming that the number of active
flavours is equal to 5. 

\subsubsection*{The non-perturbative corrections}

The theoretical cross-sections must be multiplied by the non-perturbative
coefficient $C_{\mathrm{np}}$

\begin{equation}
\left(\frac{d\sigma}{dE_{T}^{^{\gamma}}}\right)_{\mathrm{the}}=\left(\frac{d\sigma}{dE_{T}^{^{\gamma}}}\right)_{\mathrm{NLO}}*C_{\mathrm{np}}(E_{T}^{\gamma})\label{eq:as-theor}
\end{equation}

The non-perturbative corrections are divided into underlying event
and hadronization effects and $C_{\mathrm{np}}$ is calculated as
the ratio between the isolated fraction of the total prompt photon
cross section at the hadron level and the same fraction at the parton
level, obtained after turning off both (MPI) and hadronization.

The average of $C_{\mathrm{np}}$ is reported in Ref. \cite{CMS2}
as 
\begin{equation}
C_{\mathrm{np}}=0.975\pm0.006,\label{cnp used}
\end{equation}
and our estimation is given by 
\begin{equation}
C_{\mathrm{np}}=0.980\pm0.009\:(\mathrm{stat.}),\label{our cnp}
\end{equation}
 estimated using PYTHIA 8.176 with 4Cx tune parameter, both results
are close each to other but the former, which is used in our analysis,
is more relevant because it is extracted using different sets of PYTHIA
parameters.

\section{\label{sec:Extraction-of-}Extraction of $\alpha_{s}$ and averages}

\subsubsection*{The experimental data}

The measured inclusive isolated prompt photon production cross sections
$\left(d\sigma/dE_{T}^{\gamma}\right)_{\mathrm{exp}}$ are presented
as a function of the photon transverse energy $E_{T}^{\gamma}$, for
each pseudorapidity intervals.

The first 24 data points are given in Ref. \cite{ATLAS1}, were measurement
spans from $E_{T}^{\gamma}=15\:GeV$ to $E_{T}^{\gamma}=100\:GeV$
in eight $E_{T}^{\gamma}$-bins, for the |$\eta^{\gamma}|\leq0.6,\:0.6\leq|\eta^{\gamma}|<1.37\:\mathrm{and\:1.52\leq|\eta^{\gamma}|<1.81}$
regions.

32 supplementary data points are reported in Ref. \cite{ATLAS2},
in eight $E_{T}^{\gamma}$-bins between $45$ and $400\:GeV$ in the
four pseudo rapidity intervals $|\eta^{\gamma}|\leq0.6,\:0.6\leq|\eta^{\gamma}|<1.37,\:1.52\leq|\eta^{\gamma}|<1.81\:\mathrm{and\:1.81\leq|\eta^{\gamma}|<2.37}.$

Measurements are completed with 23 new data points extending significantly
the measured kinematic range to $1\:TeV$ $^{[9]}$. We have a total
of 79 experimental data points with asymmetric errors of the form:
\begin{equation}
\left[\left(d\sigma/dE_{T}^{\gamma}\right)_{\mathrm{exp}}\right]_{-\Delta_{n}}^{+\Delta_{p}}.\label{eq:sasymmerror}
\end{equation}

\begin{figure}
\centering{}\includegraphics[scale=0.71]{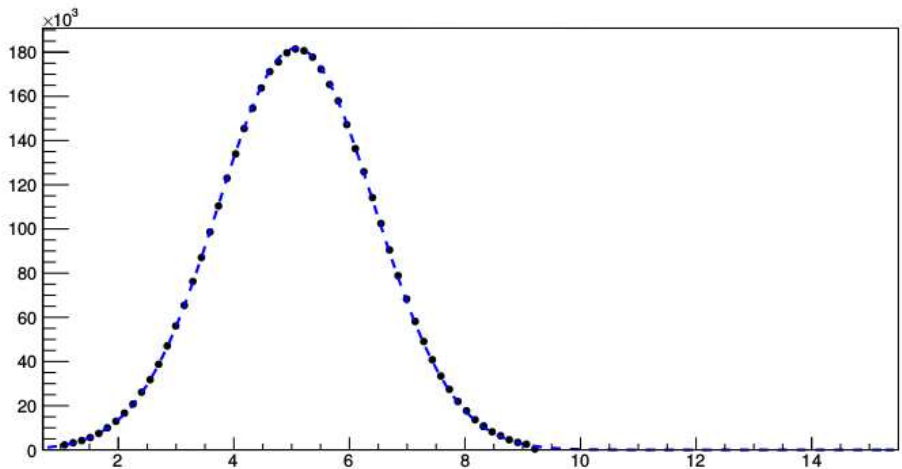}\caption{\label{fig:MC-dsEXP}An example of Gaussian distributions generated
by the Monte Carlo method corresponding to the first data point $(5.09\pm1.36)$
(nb/GeV) in the range $15<E_{T}^{\gamma}<20$ GeV and $|\eta^{\gamma}|<0.6$.
The original (asymmetrical) value is $5.24_{-1.4}^{+1.3}(\mathrm{total})\pm0.58(\mathrm{luminosity})$$^{[6]}.$The
$x$-axis represents $d\sigma/dE_{T}^{\gamma}$ in (nb/GeV) and the
$y$-axis the number of entries.}
\end{figure}

Dealing with asymmetric errors requires special care$^{[17,18]}$.
After the symmetrization following the prescriptions in Ref. \cite{AssymError},
making the results approximately symmetric and Gaussian (Fig.\ref{fig:MC-dsEXP}):

\begin{equation}
(d\sigma/dE_{T}^{\gamma})_{\mathrm{exp}}=\sigma_{\mathrm{exp}}\pm\Delta_{\mathrm{exp}},\label{eq:sexp}
\end{equation}

we propagate the experimental uncertainties by means of a series of
pseudo-experiments using Monte Carlo technique as we will see in next
section. 

\subsubsection*{The nominal values}

The value of alphas is obtained, in each of the 79 $(|\eta^{\gamma}|\mathrm{-}E_{T}^{\gamma})$-bins
of the measurement, by combining the theoretical calculations (\ref{ascurve}-\ref{cnp used})
and the $N=10^{7}$ MC generated experimental cross sections (\ref{eq:sexp}):
a set of values $\left\{ \alpha_{ik}\right\} $ are obtained for each
$(|\eta^{\gamma}|\mathrm{-}E_{T}^{\gamma})$-bin, by means of pseudo-experiments
where the experimental cross section corresponding to the bin is assumed
to be Gaussian, and then a set of cross sections $\left\{ \left(d\sigma/dE_{T}^{\gamma}\right)_{ik}\right\} $
is generated using Toy Monte Carlo techniques. Each of them is used
to extract $\alpha_{ik}$ from the theoretical curves (\ref{ascurve})
using the linear inter(extra)polation numerical method. 

The nominal value is represented by the average of the resulting sample$\left\{ \alpha_{ik}\right\} $
: 

\begin{equation}
\bar{\alpha}_{\mathrm{\mathit{i}}}=\frac{1}{N}\sum_{k=1}^{N}\alpha_{ik},\quad\mathrm{\mathit{i}\mathit{=\mathrm{1,...,79.}}}\label{eq:snom}
\end{equation}

The $\pm\sigma$ error is calculated using the CL=68\% confidence
interval for the mean. This is achieved by solving the following equation:
\begin{equation}
\intop_{\bar{\alpha}_{\mathrm{\mathit{i}}}-\Delta_{i}}^{\bar{\alpha}_{\mathrm{\mathit{i}}}+\Delta_{i}}dx\:\mathrm{pdf}_{i}(x)=\mathrm{CL}\simeq0.6827;\label{eq:CL}
\end{equation}

where $\mathrm{pdf_{i}}(x)$ is the probability density function representing
the sample $\left\{ \alpha_{ik}\right\} $.

\begin{figure}
\subfloat[$\protect\begin{array}{c}
15<E_{T}^{\gamma}<20\:\mathrm{GeV}\protect\\
\left|\eta^{\gamma}\right|<0.6
\protect\end{array}$]{\noindent \centering{}\includegraphics[scale=0.43]{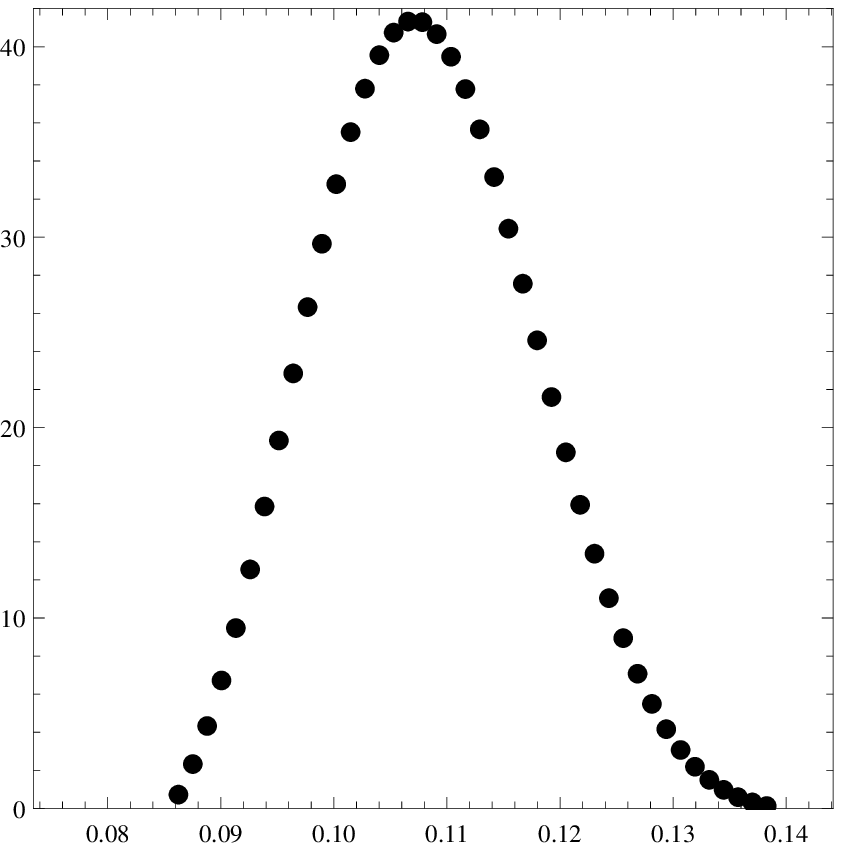}}\subfloat[$\protect\begin{array}{c}
30<E_{T}^{\gamma}<35\:\mathrm{GeV}\protect\\
\left|\eta^{\gamma}\right|<0.6
\protect\end{array}$ ]{\begin{centering}
\includegraphics[scale=0.43]{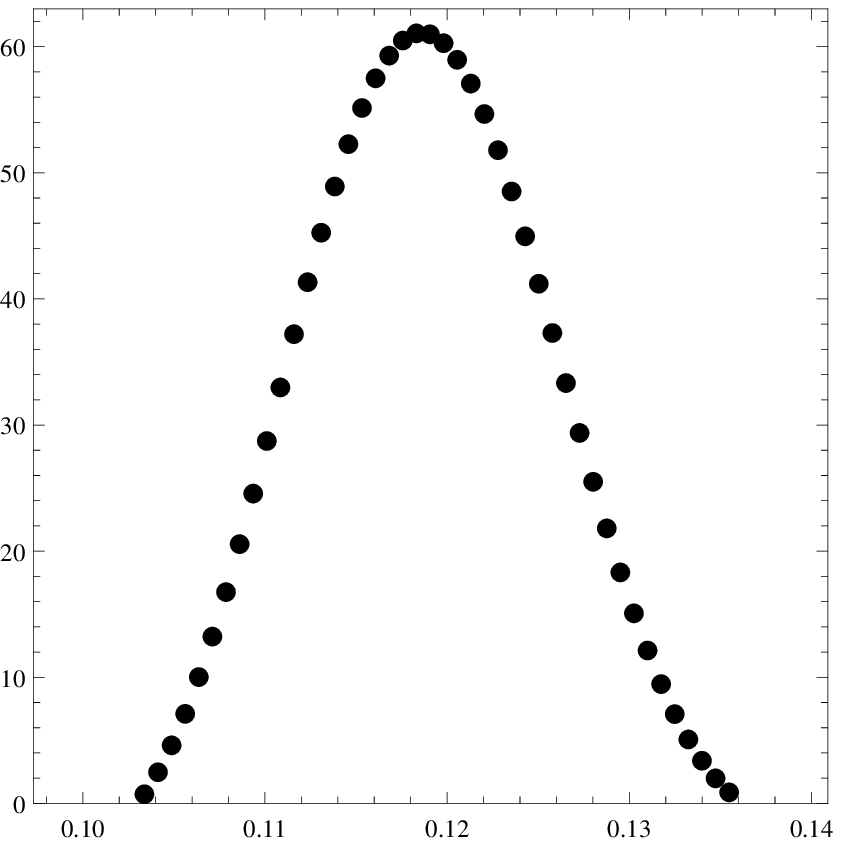}
\par\end{centering}
}\subfloat[$\protect\begin{array}{c}
45<E_{T}^{\gamma}<55\:\mathrm{GeV}\protect\\
0.6<\left|\eta^{\gamma}\right|<1.37
\protect\end{array}$ ]{\begin{centering}
\includegraphics[scale=0.43]{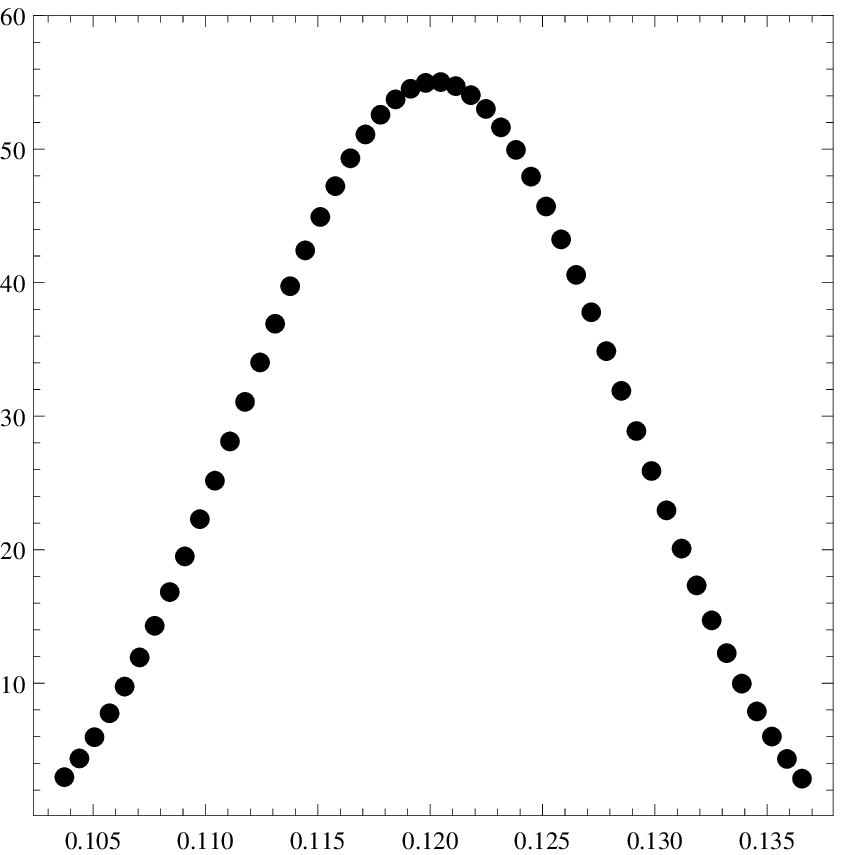}
\par\end{centering}
}

\subfloat[$\protect\begin{array}{c}
60<E_{T}^{\gamma}<100\:\mathrm{GeV}\protect\\
0.6<\left|\eta^{\gamma}\right|<1.37
\protect\end{array}$]{\begin{centering}
\includegraphics[scale=0.43]{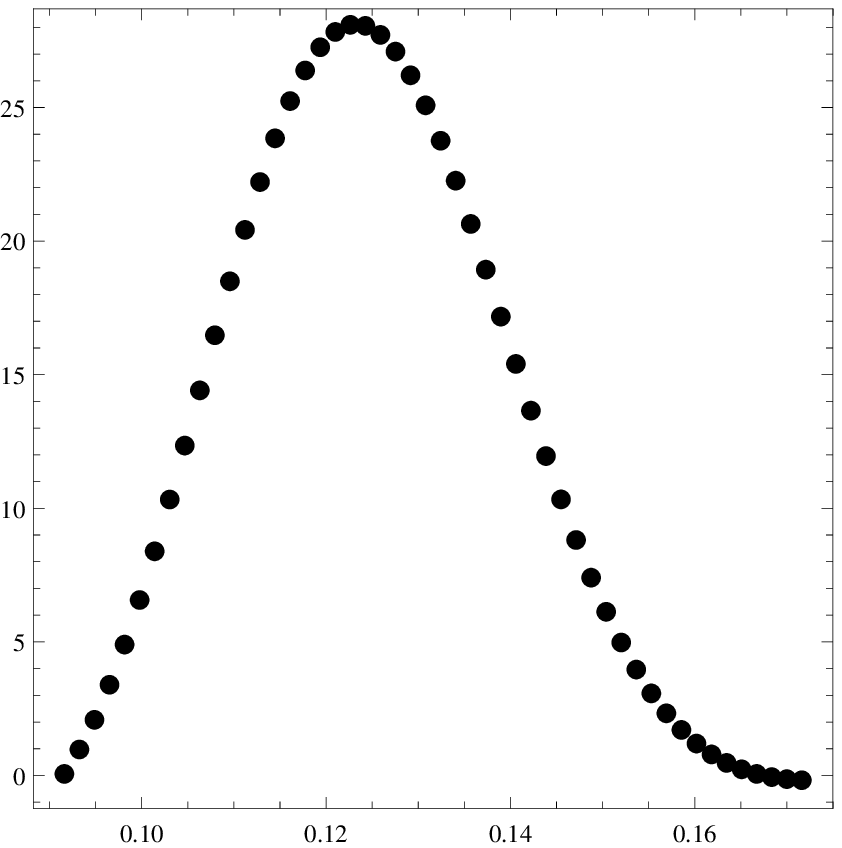}
\par\end{centering}
}\subfloat[$\protect\begin{array}{c}
200<E_{T}^{\gamma}<400\:\mathrm{GeV}\protect\\
0.6<\left|\eta^{\gamma}\right|<1.37
\protect\end{array}$]{\begin{centering}
\includegraphics[scale=0.37]{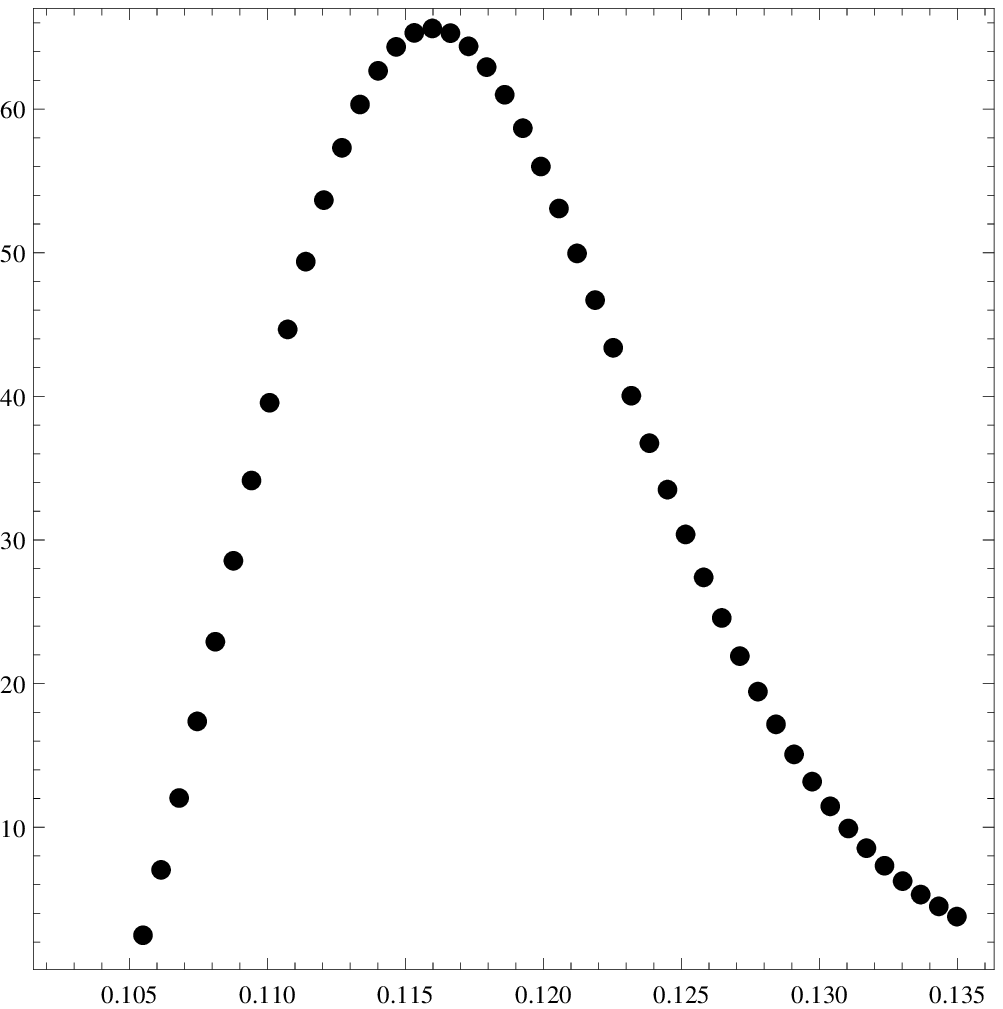}
\par\end{centering}
}\subfloat[$\protect\begin{array}{c}
500<E_{T}^{\gamma}<600\:\mathrm{GeV}\protect\\
1.52<\left|\eta^{\gamma}\right|<2.37
\protect\end{array}$]{\begin{centering}
\includegraphics[scale=0.37]{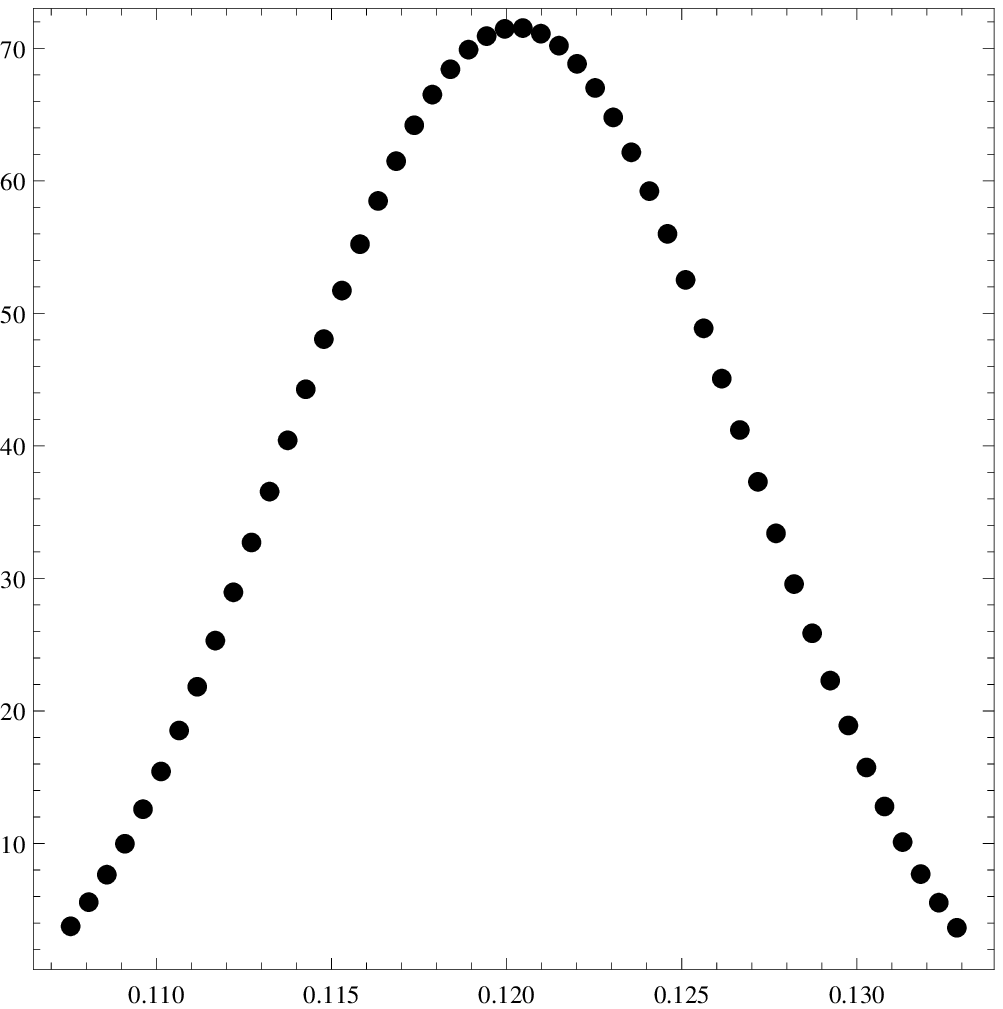}
\par\end{centering}
}

\subfloat[$\protect\begin{array}{c}
800<E_{T}^{\gamma}<1000\:\mathrm{GeV}\protect\\
\left|\eta^{\gamma}\right|<1.37
\protect\end{array}$.]{\begin{centering}
\includegraphics[scale=0.43]{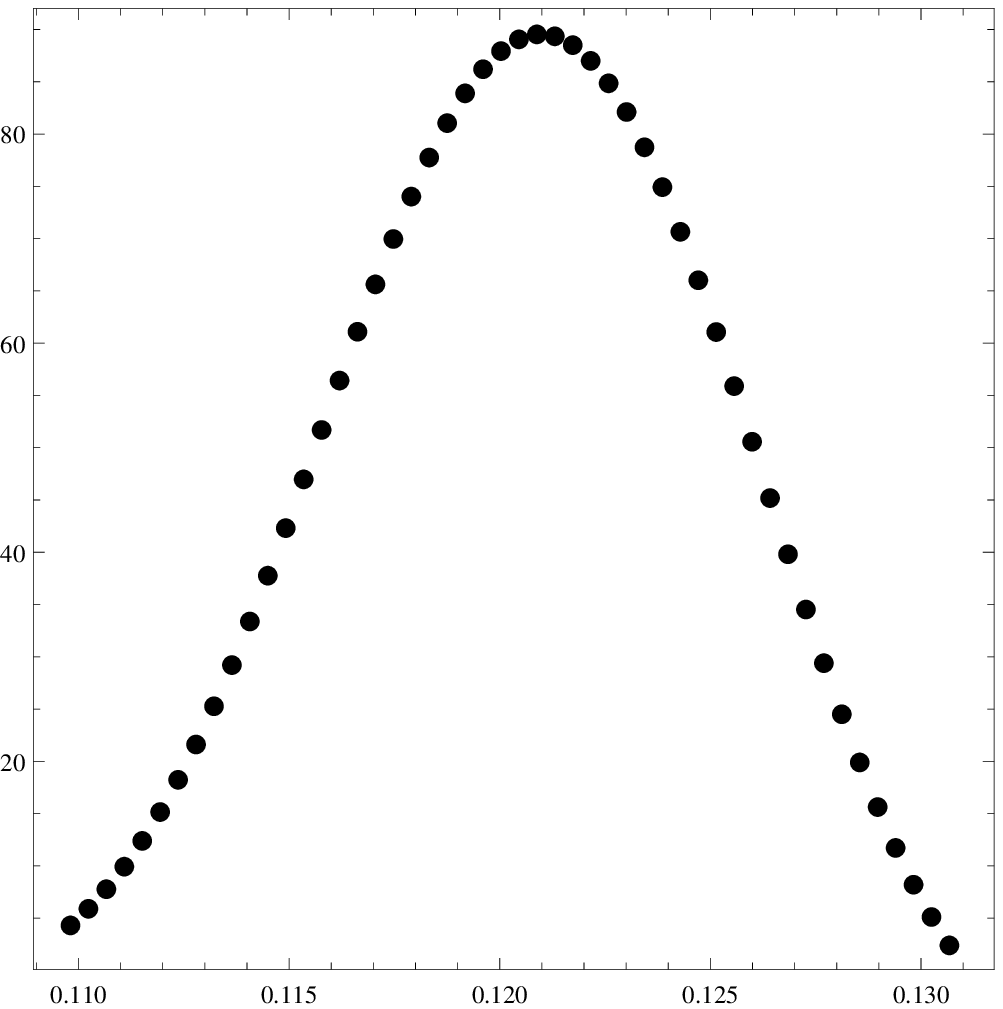}
\par\end{centering}
}

\caption{\label{fig:distributions}Examples of $\alpha_{s}$ probability distributions
constructed from data generated by pseudo-experiments in different
kinematic range. The $x$-axis represents $\alpha_{s}\left(M_{Z}^{2}\right)$
and the $y$-axis the number of entries.}
\end{figure}

The Fig.\ref{fig:distributions} shows several examples of $\alpha_{s}$distributions
extracted from pseudo-experiments in several $(|\eta^{\gamma}|\mathrm{-}E_{T}^{\gamma})$-bins.

\subsubsection*{The averaging procedure}

To obtain the average value of the strong coupling from nominal values
extracted in individual $(|\eta^{\gamma}|\mathrm{-}E_{T}^{\gamma})$
bin, we must take into account their correlations. For this purpose
we used the Best Linear Unbiased Estimate (BLUE) method$^{[19]}$
where the unbiased estimate $\hat{\alpha}$, is a linear combination
of the individual estimates,
\begin{align}
\hat{\alpha}= & \sum_{i=1}^{79}\lambda_{i}\bar{\,\alpha_{i}}\label{eq:BLUE}
\end{align}

with the constraint
\begin{equation}
\sum_{i=1}^{79}\lambda_{i}=1,\label{eq:BLUE-constraint}
\end{equation}

having the minimum variance $\sigma^{2}$:
\begin{equation}
\sigma^{2}=\sum_{ij=1}^{79}\lambda_{i}\mathrm{\,Cov_{\mathit{ij}\,}\lambda_{j},}\label{eq:BLUE-variance}
\end{equation}

$\mathrm{Cov_{\mathit{ij}}}$ represents the covariance matrix elements.

The combination of the correlated estimates requires knowledge of
the correlations between the different bins, but none of this information
is available in the published experimental papers$^{[7,8,9]}$ and
the extraction of these correlations needs full access to all the
uncertainties that contribute to the measurement of cross-sections.

Nevertheless, we have estimated the $79\times79$ ``nominal'' covariance
matrix using samples $\left\{ \alpha_{ik}\right\} $:
\begin{align}
\mathrm{Cov_{\mathit{ij}}}\simeq & \frac{1}{N}\sum_{k=1}^{N}\left(\bar{\alpha}_{\mathrm{\mathit{i}}}-\alpha_{ik}\right)\left(\bar{\alpha}_{\mathrm{\mathit{j}}}-\alpha_{jk}\right),\:\mathrm{for}\:i\neq j;\label{eq:covij}\\
\mathrm{Cov_{\mathit{ii}}}= & \Delta_{i}^{2},\nonumber 
\end{align}

this is the sample covariance matrix, an unbiased estimate of the
covariance matrix.

To be ``conservative'' the off-diagonal ``nominal'' covariance
matrix elements (\ref{eq:covij}) are multiplied by a factor in order
to maximize the variance of the combined results$^{[20]}$:
\begin{align}
\mathrm{Cov_{\mathit{ij}}} & \rightarrow f\,\mathrm{Cov_{\mathit{\mathit{ij}}},}\:\mathrm{for}\:i\neq j;\label{eq:changedCov}
\end{align}

where:
\begin{equation}
0\leq f\leq1.\label{eq:coefficient}
\end{equation}

This procedure is implemented in a software package$^{[21]}$ which
is incorporated as part of ROOT analysis framework$^{[22]}$.

For CT10wnlo PDFs and the scale choice (\ref{scales-1}), the average
procedure yields the following BLUE value: 
\begin{equation}
\alpha_{s}(M_{Z}^{2})_{\mathrm{CTEQ}}=0.1185\pm0.0010\left(\mathrm{exp}.\right).\label{alphas1}
\end{equation}

\subsubsection*{\label{subsec:The-scale-effect}The scale uncertainty}

The scale effect on the cross sections is studied in Ref. \cite{note CERN}
(Fig.10) and Ref. \cite{note ATL} (Fig.5). These bands are evaluated
by varying the three scales following the constraints:
\begin{itemize}
\item $\mu_{R}=\mu_{F}=\mu_{f}\in[\frac{1}{2}E_{T}^{\gamma},2E_{T}^{\gamma}];$
\item $\mu_{R}\text{\ensuremath{\in}}[\frac{1}{2}E_{T}^{\gamma},2E_{T}^{\gamma}];\mu_{F}=\mu_{f}=\mathrm{E_{T}^{\gamma}};$
\item $\mu_{F}\text{\ensuremath{\in}}[\frac{1}{2}E_{T}^{\gamma},2E_{T}^{\gamma}];\mu_{R}=\mu_{f}=\mathrm{E_{T}^{\gamma}};$
\item $\mu_{f}\text{\ensuremath{\in}}[\frac{1}{2}E_{T}^{\gamma},2E_{T}^{\gamma}];\mu_{R}=\mu_{F}=\mathrm{E_{T}^{\gamma}};$
\end{itemize}
The scale uncertainty on the cross section is propagated to the uncertainty
on $\alpha_{s}$ using the coefficients $\varGamma_{\mathrm{_{min}^{max}}}$
extracted from the figures cited above (see Tables \ref{tab:scale-range-coeff}):
\begin{table}
\subfloat[$\Gamma_{\mathrm{_{min}^{max}}}$ extracted from Fig. 10 of Ref. \cite{note CERN}.]{\begin{centering}
\begin{tabular}{|c||c|c|}
\hline 
$E_{T}^{\gamma}$-bin (GeV) & $\Gamma_{\mathrm{max}}$ & $\Gamma_{\mathrm{min}}$\tabularnewline
\hline 
\hline 
15-20 & 1.20 & 0.86\tabularnewline
\hline 
20-25 & 1.19 & 0.87\tabularnewline
\hline 
25-30 & 1.16 & 0.87\tabularnewline
\hline 
30-35 & 1.18 & 0.87\tabularnewline
\hline 
35-40 & 1.16 & 0.88\tabularnewline
\hline 
40-50 & 1.14 & 0.89\tabularnewline
\hline 
50-60 & 1.14 & 0.90\tabularnewline
\hline 
60-100 & 1.12 & 0.90\tabularnewline
\hline 
\end{tabular}%
\begin{tabular}{|c||c|c|}
\hline 
$E_{T}^{\gamma}$-bin (GeV) & $\Gamma_{\mathrm{max}}$ & $\Gamma_{\mathrm{min}}$\tabularnewline
\hline 
\hline 
45-55 & 1.14 & 0.89\tabularnewline
\hline 
55-70 & 1.13 & 0.90\tabularnewline
\hline 
70-85 & 1.12 & 0.90\tabularnewline
\hline 
85-100 & 1.12 & 0.91\tabularnewline
\hline 
100-125 & 1.12 & 0.91\tabularnewline
\hline 
125-150 & 1.12 & 0.90\tabularnewline
\hline 
150-200 & 1.12 & 0.90\tabularnewline
\hline 
200-400 & 1.12 & 0.90\tabularnewline
\hline 
\end{tabular}
\par\end{centering}

}

\subfloat[$\Gamma_{\mathrm{_{min}^{max}}}$ extracted from Fig. 5 of Ref. \cite{note ATL}. ]{\begin{centering}
\begin{tabular}{|c||c|c|}
\hline 
$E_{T}^{\gamma}$-bin (GeV) & $\Gamma_{\mathrm{max}}$ & $\Gamma_{\mathrm{min}}$\tabularnewline
\hline 
\hline 
100-125 & 1.059 & 0.956\tabularnewline
\hline 
125-150 & 1.065 & 0.952\tabularnewline
\hline 
150-175 & 1.068 & 0.950\tabularnewline
\hline 
175-200 & 1.070 & 0.949\tabularnewline
\hline 
200-250 & 1.068 & 0.949\tabularnewline
\hline 
250-300 & 1.063 & 0.951\tabularnewline
\hline 
300-350 & 1.054 & 0.955\tabularnewline
\hline 
350-400 & 1.043 & 0.960\tabularnewline
\hline 
400-500 & 1.038 & 0.962\tabularnewline
\hline 
500-600 & 1.048 & 0.960\tabularnewline
\hline 
600-700 & 1.061 & 0.912\tabularnewline
\hline 
700-800 & 1.076 & 0.820\tabularnewline
\hline 
800-1000 & 1.152 & 0.646\tabularnewline
\hline 
\end{tabular}%
\begin{tabular}{|c||c|c|}
\hline 
$E_{T}^{\gamma}$-bin (GeV) & $\Gamma_{\mathrm{max}}$ & $\Gamma_{\mathrm{min}}$\tabularnewline
\hline 
\hline 
100-125 & 1.076 & 0.943\tabularnewline
\hline 
125-150 & 1.084 & 0.937\tabularnewline
\hline 
150-175 & 1.091 & 0.932\tabularnewline
\hline 
175-200 & 1.095 & 0.928\tabularnewline
\hline 
200-250 & 1.098 & 0.926\tabularnewline
\hline 
250-300 & 1.100 & 0.923\tabularnewline
\hline 
300-350 & 1.085 & 0.927\tabularnewline
\hline 
350-400 & 1.067 & 0.934\tabularnewline
\hline 
400-500 & 1.067 & 0.934\tabularnewline
\hline 
500-600 & 1.055 & 0.955\tabularnewline
\hline 
\multicolumn{1}{c}{} & \multicolumn{1}{c}{} & \multicolumn{1}{c}{}\tabularnewline
\multicolumn{1}{c}{} & \multicolumn{1}{c}{} & \multicolumn{1}{c}{}\tabularnewline
\multicolumn{1}{c}{} & \multicolumn{1}{c}{} & \multicolumn{1}{c}{}\tabularnewline
\end{tabular}
\par\end{centering}
}\caption{\label{tab:scale-range-coeff}Scale ratio coefficients $\Gamma_{\mathrm{_{min}^{max}}}$in
different kinematic range. }
\end{table}

\begin{equation}
\varGamma_{\mathrm{_{min}^{max}}}=\left(d\sigma_{\mu}/d\sigma_{\mu=\mu_{R}=\mu_{F}=\mu_{f}=E}\right)_{\mathrm{_{min}^{max}}}.\label{eq:scale-effects}
\end{equation}

These values are related to MSTW2008 NLO set, but we can use them
to evaluate the scale uncertainty related to CT10 set because the
cross sections calculated with both sets are very close to each other
(see Table \ref{tab:changing-pdf-sets}).

To be ``conservative'', the extreme values \LyXZeroWidthSpace \LyXZeroWidthSpace in
each bin are considered, corresponding to the largest uncertainty.

Each theoretical curve $f_{\mathrm{bin}}$ (Eq. 2) generates two additional
curves $f_{\mathrm{_{min}^{max}}}$ by rescaling cross sections with
corresponding amounts $\varGamma_{\mathrm{_{min}^{max}}}$ and the
nominal value of the corresponding measured cross section was mapped
to the $(\bar{\alpha}_{i})_{\mathrm{min}}^{\mathrm{max}}$ .

The average procedure cited above gives two values $\alpha_{s}^{\mathrm{max}}$
and $\alpha_{s}^{\mathrm{min}}$ around the central one $0.1185$:
\begin{align}
\alpha_{s}^{\mathrm{max}} & =0.1221\label{eq:as-scale-limits}\\
\alpha_{s}^{\mathrm{min}} & =0.1175
\end{align}
 and then:
\begin{equation}
\alpha_{s}=0.1185_{-0.0010}^{+0.0036},\label{eq:as-scale}
\end{equation}

The scale uncertainty is consistent with LHC works determining $\alpha_{s}$
using jet data$^{[24]}$.

\subsubsection*{The PDF uncertainties}

\paragraph*{\uline{CTEQ eign.}}

The JETPHOX error band cross sections calculated with CT10wnlo, involving
52 member PDFs, are combined with our theoretical curves (\ref{ascurve})
to estimate PDF uncertainties.

The weighted average procedure gives: 
\begin{equation}
\bar{\alpha}_{\mathrm{CT10}}=0.1139\pm0.0028\label{eq:alpha_PDF}
\end{equation}
 with a relative error of roughly $\frac{0.0028}{0.1139}\times100=2.5\%$.
The PDF-eign. uncertainty is estimated as: 

\begin{equation}
\Delta_{\mathrm{PDF\,eig}.}^{\mathrm{CTEQ}}=\pm0.025*0.1185=\pm0.0029.\label{pdfError}
\end{equation}

This value agrees with results obtained from LHC jet data studies$^{[24]}$.\smallskip{}

\paragraph*{\uline{MSTW and CTEQ}}

We exploit informations on the cross section ratio $\varPi=d\sigma_{\mathrm{MSTW}}/d\sigma_{\mathrm{CTEQ}}$
extracted from tables 1-2 of Ref. \cite{note CERN} and Fig.4 of Ref.
\cite{ATLAS3}. We remark that there is no significant difference
between cross section values calculated with MSTW and CTEQ pdfs (see
Table\ref{tab:changing-pdf-sets}). 

The ratio $\varPi$ is used to calculate the MSTW central value of
$\alpha_{s}$:
\begin{equation}
\alpha_{s}(M_{Z}^{2})_{\mathrm{MSTW}}=0.1181\pm0.0009\left(\mathrm{exp}.\right),\label{eq:as-MSTW}
\end{equation}

then:
\begin{equation}
\alpha_{s}(M_{Z}^{2})_{\mathrm{CTEQ}}-\alpha_{s}(M_{Z}^{2})_{\mathrm{MSTW}}=0.0004.\label{eq:PDF-change}
\end{equation}

At this stage, we can write:
\begin{equation}
\alpha_{s}(M_{Z}^{2})=0.1183\pm0.0002\left(\mathrm{\mathrm{MSTW-CTEQ}}\right)\pm0.0029\left(\mathrm{CTEQ\,eig.}\right).\label{eq:as-pdf}
\end{equation}

\begin{table}
\begin{centering}
\subfloat[$\varPi$ extracted from tables 1-2 of Ref. \cite{note CERN}.]{

\begin{tabular}{|c|c|c|c|c}
\cline{1-4} 
$E_{T}^{\gamma}$-bin (GeV) & $\varPi_{\left|\eta^{\gamma}\right|<0.6}$ & $\varPi_{0.6<\left|\eta^{\gamma}\right|<1.37}$ & $\varPi_{1.52<\left|\eta^{\gamma}\right|<1.81}$ & \tabularnewline
\cline{1-4} 
15-20 & 1.027 & 1.024 & 1.023 & \tabularnewline
\cline{1-4} 
20-25 & 1.041 & 1.040 & 1.033 & \tabularnewline
\cline{1-4} 
25-30 & 1.046 & 1.050 & 1.043 & \tabularnewline
\cline{1-4} 
30-35 & 1.063 & 1.050 & 1.052 & \tabularnewline
\cline{1-4} 
35-40 & 1.036 & 1.049 & 1.048 & \tabularnewline
\cline{1-4} 
40-50 & 1.055 & 1.043 & 1.052 & \tabularnewline
\cline{1-4} 
50-60 & 1.059 & 1.062 & 1.042 & \tabularnewline
\cline{1-4} 
60-100 & 1.083 & 1.067 & 1.000 & \tabularnewline
\cline{1-4} 
\end{tabular}%
\begin{tabular}{|c|c|c|c|c|}
\hline 
$E_{T}^{\gamma}$-bin (GeV) & $\varPi_{\left|\eta^{\gamma}\right|<0.6}$ & $\varPi_{0.6<\left|\eta^{\gamma}\right|<1.37}$ & $\varPi_{1.52\leq\left|\eta^{\gamma}\right|<1.81}$ & $\varPi_{1.81\leq\left|\eta^{\gamma}\right|<2.37}$\tabularnewline
\hline 
45-55 & 1.052 & 1.050 & 1.046 & 1.041\tabularnewline
\hline 
55-70 & 1.055 & 1.053 & 1.047 & 1.041\tabularnewline
\hline 
70-85 & 1.058 & 1.055 & 1.048 & 1.040\tabularnewline
\hline 
85-100 & 1.060 & 1.056 & 1.048 & 1.037\tabularnewline
\hline 
100-125 & 1.060 & 1.054 & 1.054 & 1.037\tabularnewline
\hline 
125-150 & 1.067 & 1.062 & 1.029 & 1.036\tabularnewline
\hline 
150-200 & 1.040 & 1.031 & 1.000 & 1.000\tabularnewline
\hline 
\multicolumn{1}{|c||}{200-400} & 1.000 & 1.000 & 1.000 & 1.000\tabularnewline
\hline 
\end{tabular}

}
\par\end{centering}
\begin{centering}
\subfloat[$\varPi$ extracted from Fig.4 of Ref. \cite{ATLAS3}.]{%
\begin{tabular}{|c||c|c}
\cline{1-2} 
$E_{T}^{\gamma}$-bin (GeV) & $\varPi_{\left|\eta^{\gamma}\right|<1.37}$ & \tabularnewline
\cline{1-2} 
100-125 & 1.038 & \tabularnewline
\cline{1-2} 
125-150 & 1.050 & \tabularnewline
\cline{1-2} 
150-175 & 1.047 & \tabularnewline
\cline{1-2} 
175-200 & 1.105 & \tabularnewline
\cline{1-2} 
200-250 & 1.037 & \tabularnewline
\cline{1-2} 
250-300 & 1.032 & \tabularnewline
\cline{1-2} 
300-350 & 1.043 & \tabularnewline
\cline{1-2} 
350-400 & 1.028 & \tabularnewline
\cline{1-2} 
400-500 & 1.048 & \tabularnewline
\cline{1-2} 
500-600 & 1.020 & \tabularnewline
\cline{1-2} 
600-700 & 1.021 & \tabularnewline
\cline{1-2} 
700-800 & 1.012 & \tabularnewline
\cline{1-2} 
800-1000 & 0.991 & \tabularnewline
\cline{1-2} 
\end{tabular}%
\begin{tabular}{|c||c|}
\hline 
$E_{T}^{\gamma}$-bin (GeV) & $\varPi_{1.52\leq\left|\eta^{\gamma}\right|<2.37}$\tabularnewline
\hline 
100-125 & 1.046\tabularnewline
\hline 
125-150 & 1.018\tabularnewline
\hline 
150-175 & 1.031\tabularnewline
\hline 
175-200 & 1.001\tabularnewline
\hline 
200-250 & 0.998\tabularnewline
\hline 
250-300 & 1.019\tabularnewline
\hline 
300-350 & 1.023\tabularnewline
\hline 
350-400 & 1.036\tabularnewline
\hline 
400-500 & 0.951\tabularnewline
\hline 
500-600 & 1.010\tabularnewline
\hline 
\multicolumn{1}{c}{} & \multicolumn{1}{c}{}\tabularnewline
\multicolumn{1}{c}{} & \multicolumn{1}{c}{}\tabularnewline
\multicolumn{1}{c}{} & \multicolumn{1}{c}{}\tabularnewline
\end{tabular}

}
\par\end{centering}
\caption{\label{tab:changing-pdf-sets}The ratio of cross sections $\varPi$
corresponding to CTEQ and MSTW pdfs, in different kinematic range.}
\end{table}

\subsubsection*{The $\alpha_{s}$value }

Finally our estimation of the $\alpha_{s}(M_{Z}^{2})$, including
experimental, PDF and scale errors is:
\begin{align}
\alpha_{s}(M_{Z}^{2}) & =0.1183\pm0.0010(\mathrm{exp.})\pm_{0.0010}^{0.0036}(\mathrm{scale})\pm0.0002(\mathrm{MSTW\mathrm{-}CT10\:PDF})\pm0.0029(\mathrm{CT10\:eig.})\label{eq:as-final}\\
\alpha_{s}(M_{Z}^{2}) & =0.1183\pm0.0038.\nonumber 
\end{align}

This result is consistent with the recent PDG average world value
$0.1181\pm0.0011$$^{[2]}$, and with values extracted directly from
jet measurements at the LHC$^{[24]}$, especially with the value reported
by CMS Collaboration$^{[23]}$: $0.1185\pm0.0019\:(\mathrm{exp})\pm0.0028\:(\mathrm{PDF})\pm0.0004\:(\mathrm{NP})\pm0.0024\:(\mathrm{scale})$.

\section{\label{sec:Conclusion}Conclusion}

Using the measured inclusive isolated prompt photon production cross
sections reported by ATLAS Collaboration at $\sqrt{s}=7$ TeV combined
with Monte Carlo NLO calculations, we propose for the first time an
estimation of the strong coupling constant exploiting the prompt photon
production process, up to TeV region. Both theoretical and experimental
errors are evaluated and our result has been determined to be $\alpha_{s}\left(M_{Z}^{2}\right)=0.1183\pm0.0038\:(\mathrm{exp.,PDF},\mathrm{scale})$,
which is in good agreement with the most recent world average value
$0.1181\pm0.0011$$^{[2]}$.

It is important to note that the theoretical uncertainties are mostly
coming from terms beyond NLO order. The calculations of prompt photon
production cross sections to NNLO are necessary to overcome this deficiency,
especially they will minimize the sensitivity of the result to the
scale parameters and will improve accuracy in $\alpha_{s}$ determination.

\section{Acknowledgements}

This work was realized with the support of the FNR (Algerian Ministry
of Higher Education and Scientific Research), as part of the research
project D018 2014 0044. The authors are grateful to Olivier Pène and
Michel Fontannaz for many useful discussions on this work; and we
thank the laboratory of theoretical physics (LPT, Université Paris-Sud,
Orsay) for its warm hospitality during our visits there. In addition,
we gratefully acknowledge the UCI Computing Centre of Oran-1 University
and its Staff for delivering so effectively the computing infrastructure
essential to our work.

\end{document}